\documentclass[12pt]{iopart}
\include{graphicx}

\begin{document}

To appear in The Astronomical Journal

\title[The Collimated Jet Source in IRAS~16547-4247]{The Collimated Jet Source 
in IRAS~16547-4247: Time Variation, Possible Precession, and Upper Limits to the Proper Motions Along the Jet Axis}

\author{Luis F. Rodr\'\i guez}
\address{Centro de Radioastronom\'\i a y Astrof\'\i sica, UNAM,
Apdo. Postal 3-72, Morelia, Michoac\'an, 58089 M\'exico}
\ead{l.rodriguez@astrosmo.unam.mx}

\author{James M. Moran and Ramiro Franco-Hern\'andez\footnote{Centro de Radioastronom\'\i a y Astrof\'\i sica,
UNAM, Morelia 58089, M\'exico}}
\address{Harvard-Smithsonian Center for Astrophysics,
60 Garden Street, Cambridge, MA 02138, USA}
\ead{jmoran, rfranco@cfa.harvard.edu}

\author{Guido Garay}
\address{Departamento de Astronom\'\i a,
Universidad de Chile, Casilla 36-D, Santiago, Chile}
\ead{guido@das.uchile.cl}

\author{Kate J. Brooks}
\address{Australia Telescope National Facility, P.O. Box 76, Epping NSW 1710
Australia}
\ead{Kate.Brooks@atnf.csiro.au}


\author{Diego Mardones}
\address{Departamento de Astronom\'\i a,
Universidad de Chile, Casilla 36-D, Santiago, Chile}
\ead{mardones@das.uchile.cl}


\begin{abstract}

The triple radio source detected in association with the luminous infrared
source IRAS~16547-4247 has previously been studied with high angular resolution and high 
sensitivity with the Very Large Array (VLA) at 3.6-cm wavelength.
In this paper, we present new 3.6 cm observations taken 2.68 years
after the first epoch that allow a search for variability and proper motions, as well as 
the detection of additional faint sources in the region.
We do not detect proper motions along the axis of the outflow in the outer lobes of this source 
at a 4-$\sigma$ upper limit of $\sim$160 km s$^{-1}$. This suggests that these lobes are probably 
working surfaces where the jet is interacting with a denser medium. However, the brightest
components of the lobes show evidence of precession, at a
rate of $0\rlap.^\circ08$ yr$^{-1}$ clockwise in the plane of the sky.
It may be possible to understand the distribution of almost
all the identified sources as the result of ejecta from a precessing jet.
The core of the thermal jet shows significant variations in flux density and morphology. We compare this 
source with other jets in low and high mass young stars and suggest that the former can 
be understood as a scaled-up version of the latter.

\end{abstract}


\section{Introduction}

A successful model of low-mass star formation, based on accretion via a circumstellar disk and a 
collimated outflow in the form of jets (Shu, Adams, \& Lizano 1987), has been developed and found to 
be consistent with the observations.  An important question related to star formation is whether or not 
this model can be scaled up for the case of high-mass protostars or if other physical processes 
(i.e., stellar merging; Bonnell, Bate, \& Zinnecker 1998; Stahler, Palla, \& Ho 2000; Bally \& Zinnecker 2005) are important.
A small number of B-type young stars have been found to be associated
with collimated outflows and possibly even circumstellar disks
(see Garay \& Lizano 1999; Arce et al. 2007; Cesaroni et al. 2007). 
The source IRAS~16547-4247 is the best example of a highly-collimated outflow associated 
with an O-type protostar studied so far, and its study may reveal important  
information about the way high mass stars form.

The systemic LSR velocity of the ambient molecular cloud where IRAS~16547-4247 is embedded is 
$-$30.6 km s$^{-1}$ (Garay et al. 2007). Adopting the galactic rotation model of Brand \& Blitz (1993) and 
assuming that the one-dimensional rms velocity dispersion among molecular clouds is 7.8 km s$^{-1}$ (Stark \& Brand 1989),
we estimate a distance of 2.9$\pm$0.6 kpc for the source.
IRAS~16547-4247 has a bolometric luminosity of 6.2$\times$10$^4$ $L_\odot$, 
equivalent to that of a single O8 zero-age main-sequence star, although
it is probably a cluster for which the most massive star would have slightly lower luminosity. 
Garay et al. (2003) detected an embedded triple radio continuum source associated with the IRAS~16547-4247.
The three radio components are aligned in a northwest-southeast direction, 
with the outer lobes symmetrically separated from the central source by 
an angular distance of $\sim10{''}$, equivalent to
a physical separation in the plane of the sky of $\sim$0.14 pc. 
The positive spectral index of the central source
is consistent with that expected for a radio thermal (free-free) jet
(e.g., Anglada 1996; Rodr\'\i guez 1997),
while the spectral index of the lobes suggests a mix of thermal and nonthermal emission. 
Forster \& Caswell (1989) detected both mainline OH and H$_2$O masers
at a position close to the central continuum source.
The triple system is centered on the position of the IRAS source and is 
also coincident within measurement error with a 1.2 mm dust continuum and
molecular line emission core whose
mass is on the order of 10$^3$ $M_\odot$ (Garay et al. 2003).
Brooks et al. (2003) reported a chain of $H_2$ 2.12 $\mu$m emission 
knots that trace a collimated flow extending over 1.5 pc
that emanates from close to the central component of the triple radio
source and has a position angle very similar to that defined by the
outer lobes of the triple radio source. Most likely this extended
component traces gas ejected in the past by the central component of
the triple source. The molecular observations of
Garay et al. (2007) revealed the presence of a collimated bipolar outflow 
with lobes $\sim$0.7 pc in extent and aligned with the thermal jet 
located at the center of the core. 

In a high angular resolution study made with the VLA and ATCA,
Rodr\'\i guez et al. (2005) confirmed that the central object is a thermal radio jet, while the 
two outer lobes are most probably heavily obscured HH objects.
The thermal radio jet was resolved angularly for the first time
by these authors and found to align closely with the outer lobes. 
Several fainter sources detected in the region away
from the main outflow axis were interpreted as most probably associated
with other stars in a young cluster. Brooks et al. (2007) used ATCA observations to substantiate the 
jet nature of the central source and to detect emission at 88 GHz that most probably arises from dust associated with this source.
 
In this paper, we present new sensitive high angular-resolution Very Large Array 
observations that provide new information on the characteristics of the radio triple source in 
IRAS~16547-4247 as well as other sources in the field.
 
\section{Observations}

The new 3.6-cm wavelength observations were made using the Very Large Array (VLA) of the National 
Radio Astronomy Observatory (NRAO)\footnote{NRAO is a facility of the National Science Foundation 
operated under cooperative agreement by Associated Universities, Inc.}. These VLA radio continuum 
observations were carried out in 2006 May 31 and June 08 at the frequency of 8.46 GHz. 
The array was in the BnA configuration and an effective bandwidth of 100~MHz
was used.  The absolute amplitude 
calibrator was 1331+305 (with an adopted flux density of 5.21 Jy), and a source model provided by NRAO was used for its calibration.
The phase calibrator was 1626$-$298, with bootstrapped
flux densities of 1.684$\pm$0.003 and
1.823$\pm$0.004 Jy for the first and second epochs, respectively. The phase
center of the array was $\alpha=16^h58^m17{\rlap.}{^s}202$ and
$\delta=-42^\circ52'09{\rlap.}{''}59$ (J2000.0). The
data were edited and calibrated using
the software package Astronomical Image Processing System (AIPS) of NRAO.
To correct for amplitude and phase errors caused by the low elevation of 
the source, the data were self-calibrated in phase and amplitude.  
No significant variations were
found between the two epochs of observations (separated by only eight days) 
and the final analysis was made from the result of concatenating all data. 
The average epoch for the two data sets is 2006.42, which we use in subsequent  
analysis.
Clean maps were obtained using the task IMAGR of AIPS with the 
ROBUST parameter set to 0. The synthesized (FWHM) beam was $1\rlap.{''}18\times0\rlap.{''}60; PA = 8^\circ$. 
The noise level achieved in the final image was 23 $\mu$Jy beam$^{-1}$.

The 3.6-cm wavelength archive data was taken on 2003 September 25 
and 30 (epoch 2003.74) in very similar conditions to those of the new observations 
and the resulting beam was $1\rlap.{''}17\times0\rlap.{''}64; PA = 11^\circ$. This old data set 
is described in Rodr\'\i guez et al. (2005). These data were cross-calibrated in phase and amplitude
using the 2006.42 epoch as the model, which serves 
to globally minimize the differences in positions and flux densities of the sources. The absolute flux 
density error is estimated to be 10\%.

\section{Results and Discussion}

To allow a more reliable comparison between the two epochs, both
images were convolved to the same angular resolution
($1\rlap.{''}20\times0\rlap.{''}65; PA = 9^\circ$).
To improve the relative gain  
calibration between the two epochs we examined the relative strengths
of the 13 sources we identified in the images. It is clear that the  
ratios of the flux densities of all the sources except S-1 and N-1  
had the
same constant value of 0.91$\pm$0.02. Hence, we adjusted the flux scale  
of the first epoch by this relative gain factor. Figure 1 shows the
corrected ratio of the flux densities of all the sources. The reduced  
$\chi^2$ of the estimate of the gain constant is 0.76, which suggests
that the flux density errors may be slightly overestimated by about  
15 percent. Since the $\chi^2$ is close to unity we conclude that there
is no evidence for variability among the 11 sources, whereas the  
changes in flux densities of S-1 and the jet are highly significant.
Figure 2 shows contour images of the emission observed at
8.46 GHz in the two epochs, as well as the difference image
(2006.42 - 2003.74). The positions, flux densities, and deconvolved angular sizes of the sources 
identified in Fig. 2 are given in Table 1. 
The observed parameters for each component were determined from a
linearized least-squares fit to a Gaussian ellipsoid function using the task
JMFIT of AIPS. In Table 2
we summarize the parameters of the proper motions of all sources in the field,
derived from differences of the positions in the two epochs. Note that the position of the jet 
changes slightly between the epochs by about 6 mas in each coordinate ($\sim3\sigma$).  We do not 
attach any significance to this shift.  If we aligned the jet positions, the proper motions listed in 
Table 2 would change by about 2 mas yr$^{-1}$.  This amount is insignificant and does not change any of our conclusions.

We note that the faint features that appear marginally detected at the 4-$\sigma$ level in the images of 
the individual epochs (for example, some faint structures to the west of the central source in the 2006.42 image), do 
not appear in the difference image, since this is about $\sqrt{2}$ noisier than the individual images. 

As already discussed by Rodr\'\i guez et al. (2005), the lobes
first observed by Garay et al. (2003) break into 
several components. A more careful examination of
the 2003 map in conjunction with the 2006 map shows that there are two  
other components present in the northern lobe
(components N-4 and N-5; see Table 1 and Figure 2), as well as two additional
field sources (sources D and E, see Table 1).

In this section we will discuss the sources individually when
new information was found, and in the
following section we focus on the search for variability and proper motions. Most of the interpretation is 
in the context of a jet and bipolar outflow.  The position angle of the jet is $-16\pm1^{\circ}$, whereas the 
PA of the CO lobes is about $-6^{\circ}$ and the PA of the line joining N-1 and S-1 is $-16^{\circ}$.   We adopt 
a nominal angle for the jet outflow of $-16^{\circ}$.

\subsection{The central jet source}

An important difference with respect to the analysis of
Rodr\'\i guez et al. (2005) is that we recognize the presence of a compact source
(source D) very close to the jet source (see Fig. 2), about
$1\rlap.{''}2$ to its NW. To obtain the parameters
of the jet and of source D separately we fitted simultaneously two Gaussian ellipsoids
to this region of the image. The jet parameters for the two epochs discussed are
given in Table 1. We note that the consideration of source D in the fitting procedure
results in somewhat smaller deconvolved dimensions for the minor axis of the jet
than obtained in Rodr\'\i guez et al.~(2005).
We assume that the
opening angle of the thermal jet is the angle subtended  
by the deconvolved minor axis at a distance of one-half the
deconvolved major axis (Eisl\"offel et al. 2000).
Using the average of the deconvolved angular dimensions given in Table 1,
we then estimate the opening angle of the thermal
jet to be $\sim$15$^\circ$ (as opposed to the
value of $\sim$25$^\circ$ derived by Rodr\'\i guez et al.), 
indicating significant collimation in this massive protostar. This result suggests that jets 
from high mass young stars can be as collimated as those found in lower mass objects, where HST studies 
indicate opening angles in the range of tens of degrees in scales of tens of AU from the star (Ray et al. 2007). On 
larger physical scales the optical jets from low mass young stars are known to show recollimation, resulting 
in opening angles of a few degrees (Ray et al. 2007). This recollimation is not evident in the thermal 
radio jets, which are usually detected only close to the star.

\subsection{The northern lobe}

As noted before, we identify two additional components (N-4 and N-5) in this lobe,
indicating an almost continuous sequence of knots between the central jet and
the outermost N-1 component (see Fig. 2).
The sequence of five knots, N-1 to N-5, shows a gentle monotonic curvature that may be indicative of 
precession (see Section 5).  Four of the five knots are resolved angularly, and it is interesting to
note that all four have position angles consistent with $\sim$$160^\circ \pm 10^\circ$.
This result suggests that they are part of the northern outflow.
This orientation does not seem to be valid for component S-1, as discussed in the
next subsection.
The N-1 and N-2 continuum components are the only ones in the whole region that appear to
be associated with class I methanol masers (Voronkov et al. 2006).

\subsection{The southern lobe}

Only the component S-1 is clearly resolved, but in contrast to 
the components of the northern lobe that have 
intrinsic position angles consistent with $\sim$$160^\circ \pm 10^\circ$, 
the southern lobe's intrinsic 
position angle is $\sim$$40^\circ$.
The misalignment between the intrinsic 
position angle of S-1 and the outflow axis, suggests that
more than part of the main jet body, S-1 could be its working surface
(e.g., Chakrabarti 1988)
to the south. As we will see below, the time variability of this component
could support this interpretation. On the other hand, the presence of
an additional component (S-2) downstream suggests that our interpretation
of S-1 as a working surface may be incorrect, or that S-2 is an independent
source associated with a star. The fact that S-2 is unresolved is consistent with this interpretation. 

\subsection{Source A}

This source is clearly resolved.
It appears to be part of
a diffuse region of emission that connects with the brighter component
of the northern lobe.
Source A could actually be part of the outflow, either if the outflow
is less collimated and the ionized gas preferentially highlights regions where the flow is
interacting with dense ambient material or if source
A is part of the northern flow being deflected to the east
by interaction with a dense clump of gas. The second possibility
is interesting and it may be related to the fact that components N1 and N2
are the only ones in the region showing methanol maser emission.
It is possible that the methanol maser emission is being stimulated by the deflection of
the jet, as it interacts with N1 and N2. Finally, the position of source A
could be understood if the
outflow has precession (see discussion in Section 5). High angular 
resolution observations of a tracer of dense molecular gas are needed to advance our understanding of this source. 

\subsection{Source D}

Source D is the faint source located $1\rlap.{''}2$ to the NW of the jet,
and first identified as an independent source here. It is unresolved angularly and may probably 
trace an independent star. In the 2006.42 image, it shows a faint extension to the west.

\subsection{Source E}

Source E is a barely detected source located about $10''$ to the 
east of the core of IRAS 16547-4247. It is located outside of the 
region shown in Figure 2, and is unresolved in angular size.
A contour image of this source is shown in Figure 3. The \sl a priori \rm
probability of finding a 3.6 cm background source with a flux density 
of $\sim$0.14 mJy (the average flux density of the two epochs)
in a solid angle of $20'' \times 20''$ is only
$\sim$0.003 (Windhorst et al.~1993). We conclude that
this radio source most
probably traces a young star embedded in this region.

\section{Search for variation and proper motions}

Analysis of Figure 2 and Table 1 indicates that significant 
flux density or morphological variations
are observed only in three sources: the jet source, N-1, and S-1. 
This is not unexpected, since these are the three brightest sources in the field
and small variations in position or flux density are not evident in weaker sources
where the signal-to-noise ratio is much smaller. 
In the case of N-1 the variation is only in position since the flux density
remained constant and the flux density variations are evident only
for the jet source and S-1.
We will discuss our interpretation
of the observed variabilities and proper motions in the following subsections.

\subsection{The central jet source and source S-1}

The jet source seems to have increased its flux density by about 10\% between the
two epochs (see Table 1 and Figure 1). This difference is consistent with the fact that both thermal jets 
(Rodr\'\i guez et al. 2001; Galv\'an-Madrid, Avila, \& Rodr\'\i guez 2004) and HH knots (Rodr\'\i guez et
al. 2000) can show small but statistically significant flux density variations on scales of years or
even months. The difference image (Figure 4)  
suggests that the increase in emission comes from two discrete positions.
One increase (of about 0.8 mJy) is unresolved and associated with the central source and
we attribute it to an increase in mass loss at the core of the jet. The second
increase (of about 0.4 mJy) comes from an unresolved component
clearly displaced to the SE, and located at $\alpha(2000) = 16^h~ 58^m~ 17\rlap.^s2446 \pm
0\rlap.^s0033; \delta(2000) =  -42^\circ~ 52'~ 08\rlap.^{''}354 \pm 0\rlap.^{''}067$, 
at $1\rlap.{''}3$ from the center of the jet. 
The OH maser emission detected by Caswell (1998) is
close to the unresolved component associated with this second 
increase.  
We can think of three interpretations for this localized flux density increase.
One is that we are observing a time-variable source powered by an independent star.
There are two arguments against this interpretation: the first is that the sources
in this region in general show little variation (see Table 1) and the
second is that the source
lies exactly in the path of the outflow, suggesting
a relation with the jet. This latter fact suggests a second interpretation:
we are seeing a discrete, new ejecta from the jet. This interpretation, on its turn, has
several difficulties. The first is that, while a similar phenomenon was observed and monitored
in the jet associated with HH~80-81 (Mart\'\i, Rodr\'\i guez, \& Reipurth 1995;
1998), in this source the ejection
was clearly bipolar, while in IRAS~16547-4247 we would have to consider a monopolar
ejection. The second problem is related to the large velocity required for
the ejection to move over $1\rlap.{''}3$ in 2.68 years or less, that at a
distance of 2.9 kpc implies the unlikely velocity of $\sim10^4$~km~s$^{-1}$
or more. Finally, a third possibility is that we are observing the brightening
of the jet flow as it interacts with dense gas in its path.   
Observations of higher angular resolution are required to disentangle 
the nature of the variations seen in the jet source.
Unfortunately, the present observations are the best that it can
be done now and possibly for decades (until completion of the Square Kilometer
Array). The source can be observed, but with lower angular resolution, with
ATCA (as already done) but it cannot be observed from the latitude 
of the future e-MERLIN.  Finally, the emission is thermal and not easily detectable with VLBI arrays.

The source S-1 is the one with the largest flux density variation, with an increase of
$\sim$40\% (see Figure 1 and Table 1) between the two epochs.
We believe that this large increase is related
to our suggestion that S-1 may be a working surface of the jet, where
kinetic energy is rapidly being dissipated and changes are expected.

\subsection{Lack of Proper Motions Along the Jet Axis}

The proper motions of the prominent components N-1 and S-1
along the nominal jet axis of $-16^\circ$ are $-2 \pm 5$ and 
$-3 \pm 2$ mas yr$^{-1}$, respectively. These motions correspond
to inward motions of 28 and 42 km s$^{-1}$, respectively, and
are not statistically significant. The weighted average 
radial motion of all the components is $5.0 \pm 2.5$ mas yr$^{-1}$, corresponding
to $70 \pm 40$ km s$^{-1}$. We adopt a conservative 4-$\sigma$ upper limit on the proper motion 
of 160 km s$^{-1}$. This upper limit is not very stringent, but
certainly indicates that the IRAS~16547-4247 lobes are not moving as fast as the
knots observed in the jets associated with the massive
young stars HH~80-81 and Cep A HW2, where velocities in the plane of the sky of
$\sim$500 km s$^{-1}$ have been reported (Mart\'\i, Rodr\'\i guez, \& Reipurth 1998; Curiel et al. 2006).

\subsection{Proper Motions in the Direction Transverse to the Jet Axis}

The only components that show transverse motions to the nominal
jet direction of $-16$ degrees above the 4-$\sigma$ level are N-1 and S-1 (see Table 2). These motions 
are $-13 \pm 3$ and $-16 \pm 2$~mas~yr$^{-1}$,
corresponding to clockwise transverse motions of $180 \pm 40$ and $220 \pm 30$
km s$^{-1}$, respectively. The discovery of transverse (clockwise)
motion, with no radial motion, was unexpected.  

The total flux density of component N-1 hardly changed,
and the difference images shown in Figures 2 and 5
clearly show the negative-positive residuals characteristic of
a moving source with constant flux density. 
In contrast, the negative-positive signature of proper motion
is not present in the case of S-1 (see Figures 2 and 6) because
during the same period there was the strong brightening
previously discussed
that dominated the difference image.

\section{A Precession Model for IRAS 16547-4247}

We believe that it may be possible to understand the distribution of almost
all the identified sources as well as the clockwise
precession as the result of ejecta from a precessing jet of
stellar origin. The gently curved and antisymmetric distribution
of the radio components strongly suggests such a model. The basic idea is that the jet
itself is not observed directly (because of the lack of radial motion in the outflow). 
Rather the observed sources are the result of the jet interacting with the ambient medium. We describe this model in
simple empirical terms. We assume that the axis of symmetry is along the
N-S direction and that the position angle (measured East of North) of
the precession axis is given by
$$\theta \sim\, \theta_m \,\sin(\omega t) - \theta _0~,~~~~~~~~     (1)$$

\noindent where $\theta_m$ is the amplitude of precession, $\omega$ is the precession rate and 
$\theta_0$ is the position angle at $t=0$.  We assume that the time scale of the observable outflow 
phenomenon is much shorter than the precession period. Hence, over a short period the precession 
position angle changes linearly, $$\theta \simeq \theta_m \omega t - \theta_0 = \beta t - \theta_0~.~~~~~~~~~~~~~~(2)$$

\noindent The coordinates in the plane of the sky at the time of our observations, $t$, for 
the location  of gas ejected from the star at time $t_e$ will be
$$x(t) = r(t) \, \sin\,\theta_e~~~~~~~~~~~~~~~~~~(3)$$
$$y(t) = r(t)\cos \theta_e~~~~~~~~~~~~~~~~~~~(4)$$

\noindent where $\theta_e$ is the precession angle at the time of emission, and $r$
is the radial position of the ejecta, $r = v~(t-t_e)$.  We assume that the
ejection velocity, $v$, is constant. Since $t-t_e = r/v$ we obtain
$$x(t) = r\,\sin\,(\frac{\beta r}{v} - \theta_0)~~~~~~~~~~~~~~~~~~~(5)$$
$$ y(t) = r\,\cos\,(\frac{\beta r}{v} - \theta_0)~.~~~~~~~~~~~~~~~(6)$$

\noindent The prediction of this simple model is that $\theta$ is linearly
related to $r$, that is

$$ \theta(r) = \beta~r/v - \theta_0. ~~~~~~~~~~~~~~~~~~~~(7)$$

\noindent Figure 7 shows the position angle of each component plotted versus its radial offset from the 
center of the jet component, which we identify as the origin.  Source C deviates from the straight line 
fit to the data by about 19$^\circ$. We assume that it is not part of the jet, but a separate entity from the 
jet, perhaps evidence of another nearby star.
The straight-line fit of equation 7 to the remaining 9 sources gives parameters,
$\theta_0 = -43^\circ \pm 4^\circ$ (the current PA of the jet), and
$\alpha = \beta/v = 2.3 \pm 0.4$ degrees/arcsecond. For a distance of 2.9 kpc the value of $\alpha$ in 
the source frame is 9.2 $\times 10^{-19}$ radians/cm or 2.9 radians/pc.  Since the jet seems to be 
precessing linearly with time, we assume that the precession period is much greater than the time scale of the 
flow. Note that we cannot solve separately for the
precession rate and the ejection velocity. The shape of the jet at the current epoch, as determined by 
equations 5 and 6, is shown in Figure 8.

As noted before only two sources have significant proper motions, N-1 and S-1. Their motions are
both in the transverse direction to the direction of the jet, whereas their motions along the jet are 
insignificant (see Table 2). Hence, we assume that
the sources are caused by the interaction of the jet and clumps of ambient material as the jet sweeps across the 
clumps. In this model, the clumps will have apparent transverse velocities given by

$$ v_T = r \beta = r \alpha v~, ~~~~~~~~~~~~~~~~~~~~~(8)$$

\noindent
which, with the observed value of $\alpha$, becomes

$$ v_T = 0.42 v (r/10")~. ~~~~~~~~~~~~~~~~~~~~~~~~~(9)$$

\noindent
The proper motion measurements of all the sources are listed in Table 2.
We believe that the apparent motions of
source C could be due to the presence of an additional, time-variable component
to the NW of the main component of source C (see Fig. 2).
Source N-4 shows proper motions that are slightly below 4-$\sigma$ (see
Table 2). However, these marginal proper motions are against the
expected flow of the jet and, if real, are probably due to a ``Christmas tree''
effect more that to a real motion.

The velocity vectors of S-1 and N-1 are nearly transverse to the jet direction, and agree in direction with 
the sense of precession of the
jet (see Fig. 8). These velocities are 180 and 220~km~s$^{-1}$, respectively.  Since N-1 and S-1 both have about 
the same radial distance, we cannot confirm the expected linear trend in transverse velocity with radius predicted 
by eqn. 8. However, with these two data points we estimate from eqn. 9 that the ejection velocity is 
$490~\pm~80$~km~s$^{-1}$. If we assume that we are indeed dealing with a working surface, the lack of 
proper motions ($\leq$160~km~s$^{-1}$) allow us to estimate the ratio between the density 
of the medium and the density of the jet.  Using the 
formulation of 
Raga, Rodr\'\i guez, \& Cant\'o (1997), 
the ratio of ambient medium density, $\rho_a$, to jet density, $\rho_j$, is 

$${{\rho_a} \over {\rho_j}} = \biggl({{v_j} \over {v_{ws}}} - 1 \biggr)^2,~~~~~~~~~~(10)$$

\noindent where $v_j$ and $v_{ws}$ are the jet and working surface velocities, respectively.
For a jet velocity of 490~km~s$^{-1}$ and an upper limit of 160~km~s$^{-1}$ for the
velocity
of the working surface, we obtain $\rho_a/\rho_j \geq$ 4. 

The precession rate of the jet is $\beta = \alpha v$. Hence with  
$\alpha = 2.3\pm0.4$ degree/arcsecond, and v = $490\pm80$~km~s$^{-1}$, $\beta = 0.080\pm~0.02$ degrees/year. 
Note that the measurements  
of $\alpha$ and v are both 6-$\sigma$ results, but the measurement of $\beta$ is  
a 4-$\sigma$ result.
The precession period of the jet is $T=2 \pi \theta_m/(\alpha v)$, which for a precession opening angle of 
30$^\circ$, would be about 5500 years. Since the range in PA among the components is about 25$^{\circ}$ 
(see Fig. 7), the jet travel time to outer components is about 300 years.

We might expect that the sources excited by the passing jet would persist after the jet moves on.  However, there 
is no evidence (see Fig. 8) of structure in the transverse direction in any of the sources.  This absence of ``trails''  
suggests that the decay times for emission must be less than $\sim~100$ years, which would produce an extension of 1.5$''$ 
at the radius of sources N-1 and S-1. The recombination time for an H~II region with electron density $n_e$
is estimated as 
%
%
$$\left(\frac{t}{\mbox{yr}}\right)=1.2\times10^5\left(\frac{n_e}{\mbox{cm}^{-3}}\right)^{-1}.~~~~~~(11)$$

\noindent The spectral index of N-1 indicates it is probably a thermal
optically thin HII region (Rodriguez et~al. 2005). Using the values for the
flux density and size in Table~1 we estimate a emission measure of $1.8\times10^6$ cm$^{-6}$ pc$^{-1}$ and the
electron density as $n_e=1.1\times10^4$ cm$^{-3}$.  Using this value for
the electron density in  eqn.~(10) we find a recombination time
for the N-1 component to be about 11 years. This time is an order of
magnitude smaller than that required for the jet to leave an observable transverse trail.  

For the S-1 source the emission may have a nonthermal component 
(Rodriguez et~al 2005). The density of relativistic
electrons, $n_{er}$, and magnetic field, $B$, can be estimated using
eqns.~(2) and (3) from Garay et~al. (1996)
%
$$\left(\frac{B}{\mbox{mG}}\right)=0.50
\left(\frac{S_{\nu}}{\mbox{mJy}}\right)^{2/7}
\left(\frac{\theta_{s}}{\mbox{arcsec}}\right)^{-6/7}
\left(\frac{\nu}{10~\mbox{GHz}}\right)^{1/7}
\left(\frac{D}{\mbox{kpc}}\right)^{-2/7}
\left(\log\frac{E_{max}}{E_{min}}\right)^{2/7},~~~~~~~~~~~~~~~~~~~~~~~~~(12)$$
%
$$\left(\frac{n_{er}E_{min}}{10^{-9}~\mbox{ergs cm}^{-3}}\right)=3.58
\left(\frac{S_{\nu}}{\mbox{mJy}}\right)
\left(\frac{\theta_{s}}{\mbox{arcsec}}\right)^{-3}
\left(\frac{\nu}{10~\mbox{GHz}}\right)^{1/2}
\left(\frac{D}{\mbox{kpc}}\right)^{-1}
\left(\frac{B}{\mbox{mG}}\right)^{2/7},~~~~~~~~~~~~~~~~~~~~~~~~~(13)$$
%
where $E_{max}$ and $E_{min}$ are the maximum and minimum energies of
the relativistic electrons that we assume are $10^{11}$ and $10^{6}$ eV
respectively. For the distance $D=2.9$ kpc and $\theta_s=0.76$ arcsec, we get  $B=1.4$ mG and a density of 
relativistic electrons of $n_{er}=4.6\times10^{-3}$
cm$^{-3}$. The decay time is given by  Krolick (1999) as $t_{d}=1.3\times10^{12}\nu^{-1/2}B^{-3/2}$ sec with 
$\nu$ in Hz and $B$ in G, or $t_{d}(yr) = 0.5B^{-3/2}$ for our frequency.  Using the value derived above for $B$ 
we get $t_{d}=9\times10^4$ yr, which is much too long.  
A magnetic field of 30 mG would be required to achieve a synchrotron decay time of 100 years.   

Franco-Hern\'andez et al. (2008,
in preparation) have discovered two linear structures traced by water vapor masers on the mas scale. 
Their g2 group of maser has a PA of about $-30^\circ$, which is 13$^\circ$ 
away from the current PA of the proposed precessing source 
($-43^\circ$). Another group of masers (g1) has a position angle of  about 50$^\circ$.  If these masers 
were in a disk, their pole PA would
be $-40^\circ$, close to the current precession angle. Hence these masers may be associated with the precessing excitation source.

The larger scale CO lobes are offset from the jet by about 15{''} along a PA of 
about $-6^{\circ}$. The line of sight velocity with
respect to the ambient cloud is about $\pm$30 km s$^{-1}$. These lobes could also be associated 
with the precessing source. The expected PA for emission at 15{''} from eqn. (7) would be about $-8^{\circ}$, close to 
the observed value. If the molecular flow picks up the full jet velocity of 490 km s$^{-1}$, then the inclination 
would be about 90 -- arctan (30/490), or about 86$^\circ$, very close to the value inferred from the analysis of the 
velocity structure of the CO lobes (Garay et al 2007). More realistically, the molecular flow velocity will be 
smaller, implying a smaller inclination angle. 

The precession model nicely accounts for the fact that the line from N-1 to S-1 does not intercept the
known jet.  However, if 
the angle of the driving precessing source is currently $-43^\circ$, it is significantly discrepant with the 
PA of $-16$ of the known radio jet. A possible explanation for this discrepancy might be that the central source is really 
a binary stellar system. Source I would be a non-precessing source associated with the known jet, which is related to 
the large scale CO outflow with about the same PA. Source II would be a precessing source which drives the continuum 
thermal sources and may be associated with the water masers and also the large scale CO outflow.

\section{Putting the IRAS~16547-4247 Jet in Context}

How does the IRAS~16547-4247 jet compare with other jets
found in regions of star formation? Anglada et al. (1992) have compared
the centimeter radio luminosity (taken to be proportional
to the flux density times the distance squared, $S_\nu d^2$) of thermal jets associated with
low mass young stars with the momentum rate ($\dot P$) in the associated molecular outflow.
For 16 sources studied, they conclude that these low mass objects can be fitted with a power law 
of the form $\dot P = 10^{-2.6} (S_\nu d^2)^{1.1}$, where
$\dot P$ is in $M_\odot$ yr$^{-1}$ km s$^{-1}$, $S_\nu$
is in mJy and $d$ is in kpc. In Figure 9, we plot the data used
by Anglada et al. (1992). They interpret the fit to be in agreement with a simple model in which the 
observed ionization is produced by shocks (Curiel, Cant\'o, \& Rodr\'\i guez 1987; Curiel et al. 1989), where 
about 10\% of the energy in the jet is thermalized. We have plotted in the same figure, the three best studied 
cases of thermal jets associated with massive young stars: IRAS~16547-4247,
HH~80-81, and Cep A HW2, using the data listed in
Table 3. Remarkably, these three data points fall reasonably well on the
Anglada et al.~correlation. We believe that this agreement suggests that the
jets associated with massive young stars may be a scaled-up version of the 
phenomenon seen in low mass young stars, although 
a firmer conclusion
requires the study of a larger sample of objects than now available. Note that the intrinsic radio luminosity of the 
IRAS~16547-4247 jet is about $2~\times~ 10^3$ larger than the typical radio luminosity of jets associated with 
low mass stars, and that it is the most luminous case known. 

\section{Conclusions}

Our main conclusions follow:

1) We present new, sensitive 3.6-cm wavelength VLA observations 
of the multiple radio source associated with the luminous infrared
source IRAS~16547-4247, the most massive example known of a thermal
jet found in association with a forming star. The main
purposes of these new observations were to
search for variability and proper motions as well as to 
detect additional faint sources in the region.
We detected four new components (N-4, N-5, D, and E) in the region;

2) We do not detect proper motions along the
axis of the flow in the outer lobes of this source at a 
4-$\sigma$ upper limit of $\sim$160~km~s$^{-1}$, suggesting that if
these lobes are the working surfaces of the jets and the
jet velocity is $\sim$490km~s$^{-1}$, the ambient medium is at
least four times as dense as the jet;

3) The brightest components of the lobes, sources N-1 and S-1,
show evidence of clockwise precession, at a
rate of $0\rlap.^\circ08$ yr$^{-1}$ in the plane of the sky;

4) A precessing model can account for the antisymmetric distribution of most of the sources in the field, as well as 
for the evidence of precession in
sources N-1 and S-1;

5) The thermal jet at the core of the region shows significant variations in flux density and morphology but 
our angular resolution is insufficient to reach a clear conclusion on what produces
these changes; and

6) The correlation found by Anglada et al.~(1992) for outflows and jets
in low mass stars extends to the handful of massive  forming stars
known. This result suggests that the
jets associated with massive young stars are a scaled-up version of the phenomenon
seen in low mass young stars, although a firmer conclusion
requires the study of a larger sample of jets that now available.

\ack

We thank an anonymous referee for valuable comments.
LFR acknowledges support from grant CB0702172\_3 of COECyT,
Michoac\'an, M\'exico.
DM and GG acknowledge 
support from the Chilean {Centro de Astrof\'\i sica} FONDAP 15010003.

\section*{References}
\begin{harvard}

\item[] Anglada, G., Rodr\'\i guez, L. F., Cant\'o, J., 
Estalella, R., \& Torrelles, J. M. 1992, ApJ, 395, 494

\item[] Anglada, G. 1996, in ASP Conf. Ser. 93, Radio 
Emission from the Stars and the Sun, ed. A. R. Taylor \& J. M. Paredes 
(San Francisco: ASP), 3

\item[] Arce, H. G., Shepherd, D., Gueth, F., Lee, C.-F., 
Bachiller, R., Rosen, A., \& Beuther, H.
2007, in Protostars and Planets V, ed. B. Reipurth, D. Jewitt, \& K. Keil 
(Tucson: Univ. Arizona Press), 245 


\item[] Bally, J. \& Zinnecker, H. 2005, AJ, 129, 2281 

\item[] Bonnell, I.~A., Bate, M.~R.,
\& Zinnecker, H. 1998, MNRAS, 298, 93 

\item[] Brand, J. \& Blitz, L. 1993, A\&A, 275, 67

\item[] Brooks, K. J., Garay, G., Mardones, D., \&
Bronfman, L.  2003, ApJ, 594, L131

\item[] Brooks, K. J., Garay, G., Voronkov, M., \& Rodr\'\i guez, L. F.
2007, ApJ, 669, 459

\item[] Caswell, J. L. 1998, MNRAS, 297, 215

\item[] Cesaroni, R., Galli, D., Lodato, G., Walmsley, C. M., \& Zhang, Q. 
2007, in Protostars and Planets V, ed. B. Reipurth, D. Jewitt, \& K. Keil 
(Tucson: Univ. Arizona Press), 197

\item[] Chakrabarti, S.~K.\ 1988, 
MNRAS, 235, 33 


\item[] Curiel, S., Cant\'o, J., \& Rodr\'\i guez, L. F. 
1987, RevMexA\&A, 14, 595

\item[] Curiel, S., Rodr\'\i guez, L. F., 
Bohigas, J., Roth, M., Cant\'o, J., \& Torrelles, J. M.
1989, ApL\&C, 27, 299


\item[] Curiel, S., et al. 2006, ApJ, 638, 878

\item[] Eisl\"offel, J., Mundt, R., Ray, T. P., \& Rodr\'\i guez, L. F. 2000, 
in Protostars and Planets IV, ed. V. Mannings, A. P. Boss, \& S. S. Russell 
(Tucson: Univ. Arizona Press), 815

\item[] Forster J. R., \& Caswell J. L., 1989, A\&A, 213, 339

\item[] Galv\'an-Madrid, R., Avila, R., \& Rodr\'\i guez, L. F.
2004, RevMexA\&A, 40, 31

\item[] Garay, G., Ramirez, S., Rodr\'\i guez, L. F.,
Curiel, S., \& Torrelles, J. M. 1996, ApJ, 459, 193

\item[] Garay, G., \& Lizano, S. 1999, PASP, 111, 1049   

\item[] Garay, G., Brooks, K. J., Mardones, D., \& Norris, R. P.
2003, ApJ, 587, 739

\item[] Garay, G., Mardones, D., Bronfman, L., Brooks, K.J.  Rodr\'\i guez,
L.F., G\"usten, R., Nyman, L-{\AA}, Franco-Hern\'andez, R., \& Moran, J.M.  2007, 
A\&A 463, 217



\item[] Krolik, J. H. 1999, Active galactic nuclei: from the central black hole to the galactic 
environment (Princeton, N.J.: Princeton University Press)

\item[] Mart\'\i. J., Rodr\'\i guez, L. F., \& Reipurth, B. 1993,
ApJ, 416, 208

\item[] Mart\'\i. J., Rodr\'\i guez, L. F., \& Reipurth, B. 1995,
ApJ, 449, 184 

\item[] Mart\'\i. J., Rodr\'\i guez, L. F., \& Reipurth, B. 1998, 
ApJ, 502, 337


\item[] Narayanan, G., \& Walker, C. F. 1996, ApJ, 466, 844

\item[] Raga, A. C., Rodr\'\i guez, L. F., \& Cant\'o, J. 1997, RMxF, 43, 825 

\item[] Ray, T., Dougados, C., Bacciotti, F., Eisl\"offel, J., \& Chrysostomou, A. 
2007, in Protostars and Planets V, ed. B. Reipurth, D. Jewitt, \& K. Keil
(Tucson: Univ. Arizona Press), 231



\item[] Rodr\'\i guez, L. F., Curiel, S., Moran, J. M., Mirabel, I. F., 
Roth, M. \& Garay, G. 1989, ApJ, 346, L85

\item[] Rodr\'\i guez, L. F. 1997, in 
Herbig-Haro Flows and the Birth of Low Mass Stars, proceedings
of IAU Symp. No. 182, eds. B. Reipurth \& C. Bertout, p. 83 
(Dordrecht: Kluwer)

\item[] Rodr\'\i guez, L. F., Delgado-Arellano, V. G.,
G\'omez, Y., Reipurth, B., Torrelles, J. M., Noriega-Crespo, A., Raga, A. C., \&
Cant\'o, J. 2000, AJ, 119, 882

\item[] Rodr\'\i guez, L. F., Torrelles, J. M., Anglada, G., \& Mart\'\i,
J. 2001, RevMexA\&A, 37, 95


\item[] Rodr\'\i guez, L. F., Garay, G., Brooks, K. J., \& 
Mardones, D. 2005, ApJ, 626, 953

\item[] Shu, F. H., Adams, F. C., \& Lizano, S. 1987, 
ARA\&A, 25, 23

\item[] Stahler, S.~W., Palla, F., \& Ho, P.~T.~P. in Protostars and
Planets IV (eds. Mannings, V., Boss, A.P. \& Russell, S. S.), 327
(University of Arizona Press, Tucson, 2000)

\item[] Stark, A. A., \& Brand, J. 1989, ApJ, 339, 763

\item[] Voronkov, M. A., Brooks, K. J., Sobolev, A. M., Ellingsen, S. P., 
Ostrovskii, A. B. \& Caswell, J. L. 2006, MNRAS, 373, 411


\item[] Windhorst, R. A., Fomalont, E. B., Partridge, R. B., \& 
Lowenthal, J. D. 1993, ApJ, 405, 498 

\item[] Yamashita, T., Suzuki, H., Kaifu, N., Tamura, M., Mountain, C. M., \& 
Moore, T. J. T. 1989, ApJ, 347, 894

\end{harvard}

\clearpage

\begin{figure}
\begin{center}
\includegraphics[scale=0.7, angle=0]{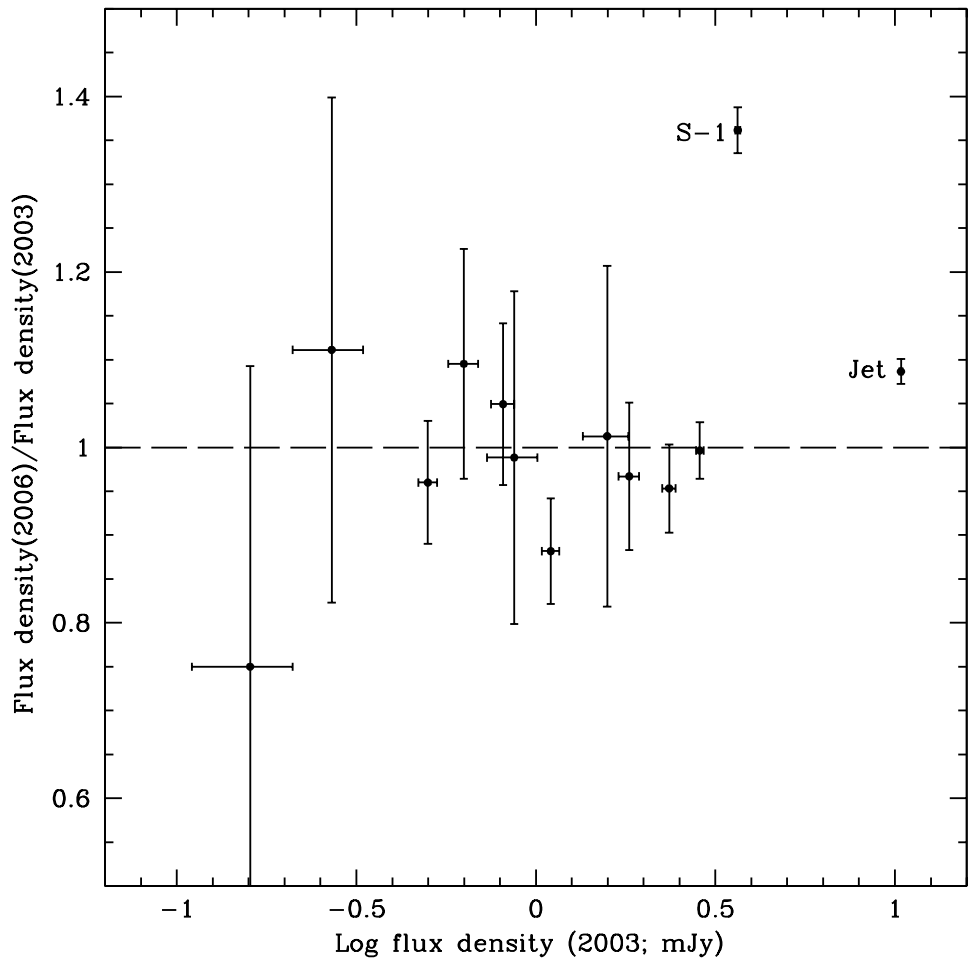}
\end{center}
\caption{Ratio of 2006 to 2003 flux densities versus the
logarithm of the 2003 flux density, given in mJy. Only the jet and source
S-1 show clear evidence of variability between the two epochs.
\label{fig1}}
\end{figure}

\begin{figure}
\begin{center}
\includegraphics[scale=0.45, angle=90]{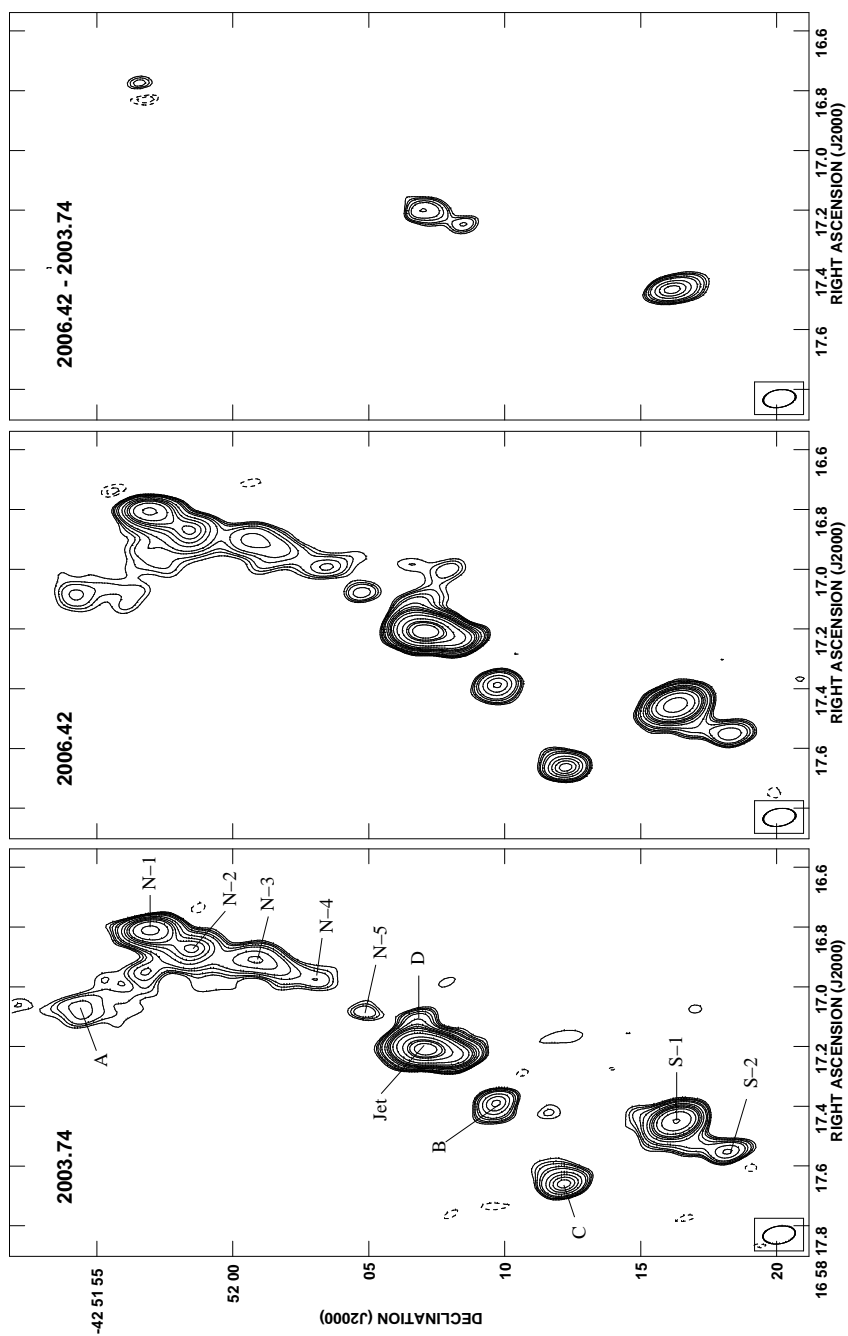}
\end{center}
\caption{VLA contour images at 8.46 GHz towards
IRAS~16547-4247 for epochs 2003.74 (left) and 2006.42 (center),
as well as the difference image (2006.42 - 2003.74).
Contours are -25, -20, -15, -10, -8, -6, -5, -4, 4, 5,
6, 8, 10, 15, 20, 25, 30, 40, 60, 100, 140, 160, and 200 times
27 $\mu$Jy beam$^{-1}$ for the 2003.74 and 2006.42 images
(the average value of the rms noises of the two images)
and 38 $\mu$Jy beam$^{-1}$ for the difference image
(the rms noise of this last image).
The half power contour of the synthesized beams
($1\rlap.{''}20 \times 0\rlap.{''}65$; PA = $9^\circ$)
is shown in the bottom left corner of each panel.
The individual sources are identified
in the 2003.74 image.
\label{fig2}}
\end{figure}

\clearpage

\begin{figure}
\begin{center}
\includegraphics[scale=0.6, angle=0]{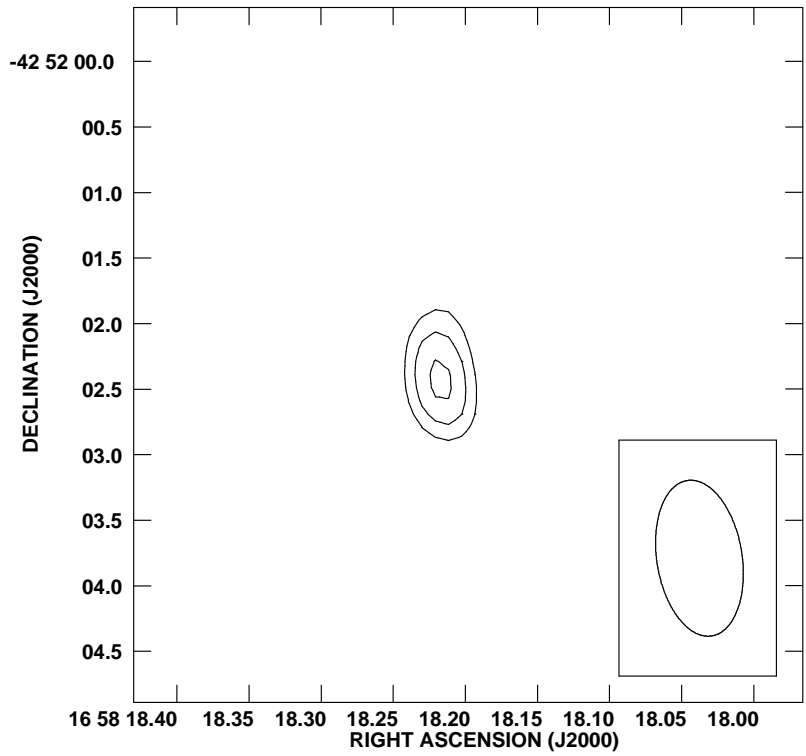}
\end{center}
\caption{VLA contour images at 8.46 GHz of
the source E, made from the average of both epochs
(2003.74 and 2006.42).
Contours are -4, 4, 5, and 6 times
18 $\mu$Jy beam$^{-1}$,
the rms noise of the image.
The half power contour of the synthesized beam
is as in Fig. 2.
\label{fig3}}
\end{figure}

\clearpage

\begin{figure}
\begin{center}
\includegraphics[scale=0.35, angle=90]{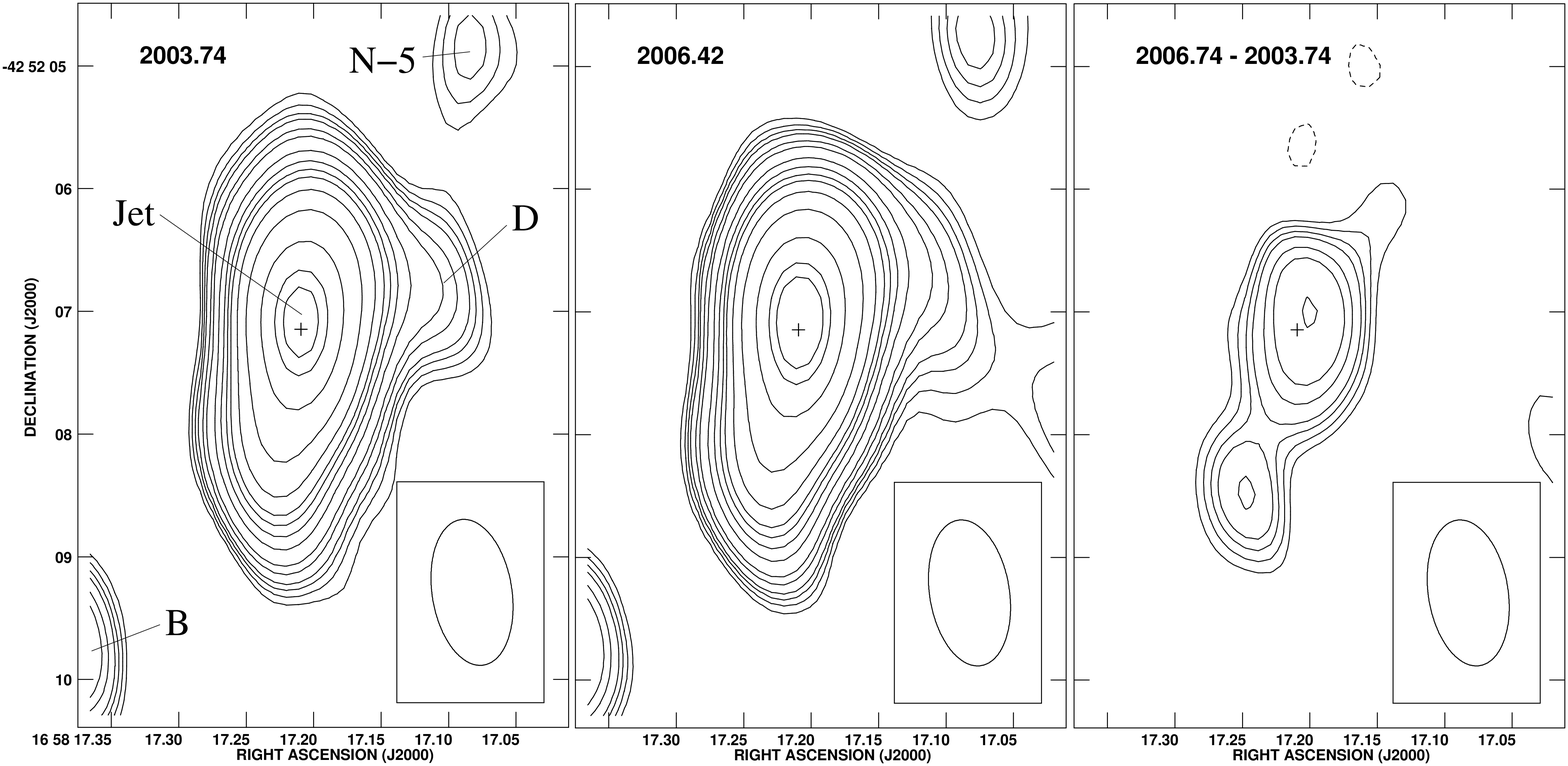}
\end{center}
\caption{VLA contour images at 8.46 GHz towards
the jet in
IRAS~16547-4247 for epochs 2003.74 (left) and 2006.42 (center),
as well as the difference image (2006.42 - 2003.74).
Contours are -25, -20, -15, -10, -8, -6, -5, -4, 4, 5,
6, 8, 10, 15, 20, 25, 30, 40, 60, 100, 140, 160, and 200 times
27 $\mu$Jy beam$^{-1}$ for the 2003.74 and 2006.42 images
and 38 $\mu$Jy beam$^{-1}$ for the difference image.
The half power contour of the synthesized beams
($1\rlap.{''}20 \times 0\rlap.{''}65$; PA = $9^\circ$)
is shown in the bottom right corner.
The cross marks the peak position of 
the jet as determined from the average of the two positions given
in Table 1.
Individual sources are identified
in the 2003.74 image.
\label{fig4}}
\end{figure}

\clearpage

\begin{figure}
\begin{center}
\includegraphics[scale=0.35, angle=90]{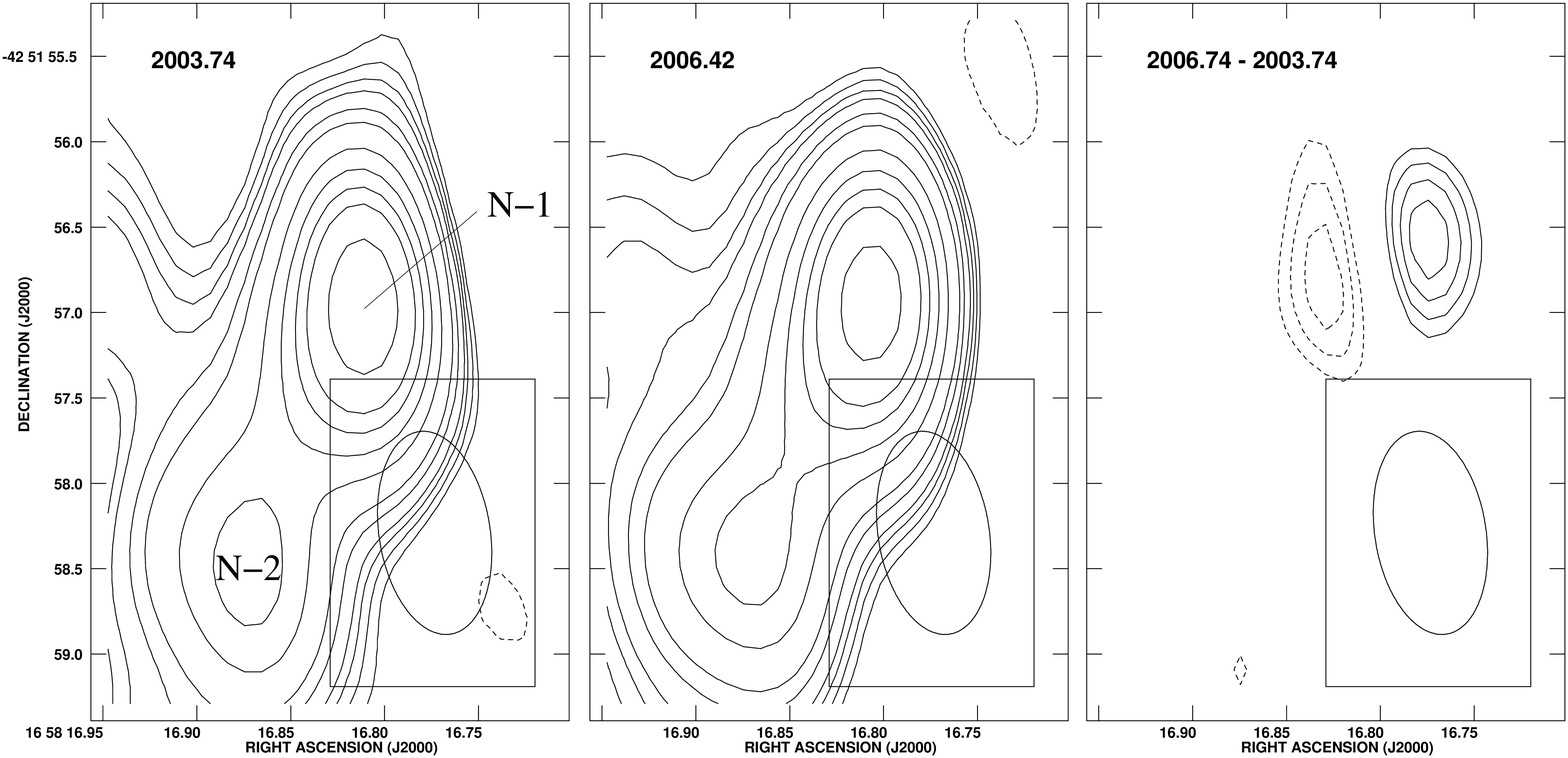}
\end{center}
\caption{VLA contour images at 8.46 GHz towards
the component N-1 in
IRAS~16547-4247 for epochs 2003.74 (left) and 2006.42 (center),
as well as the difference image (2006.42 - 2003.74).
Contours are -25, -20, -15, -10, -8, -6, -5, -4, 4, 5,
6, 8, 10, 15, 20, 25, 30, 40, 60, 100, 140, 160, and 200 times
27 $\mu$Jy beam$^{-1}$ for the 2003.74 and 2006.42 images
and 38 $\mu$Jy beam$^{-1}$ for the difference image.
The half power contour of the synthesized beams
($1\rlap.{''}20 \times 0\rlap.{''}65$; PA = $9^\circ$)
is shown in the bottom right corner.
Individual sources are identified
in the 2003.74 image.
\label{fig5}}
\end{figure}

\clearpage

\begin{figure}
\begin{center}
\includegraphics[scale=0.35, angle=90]{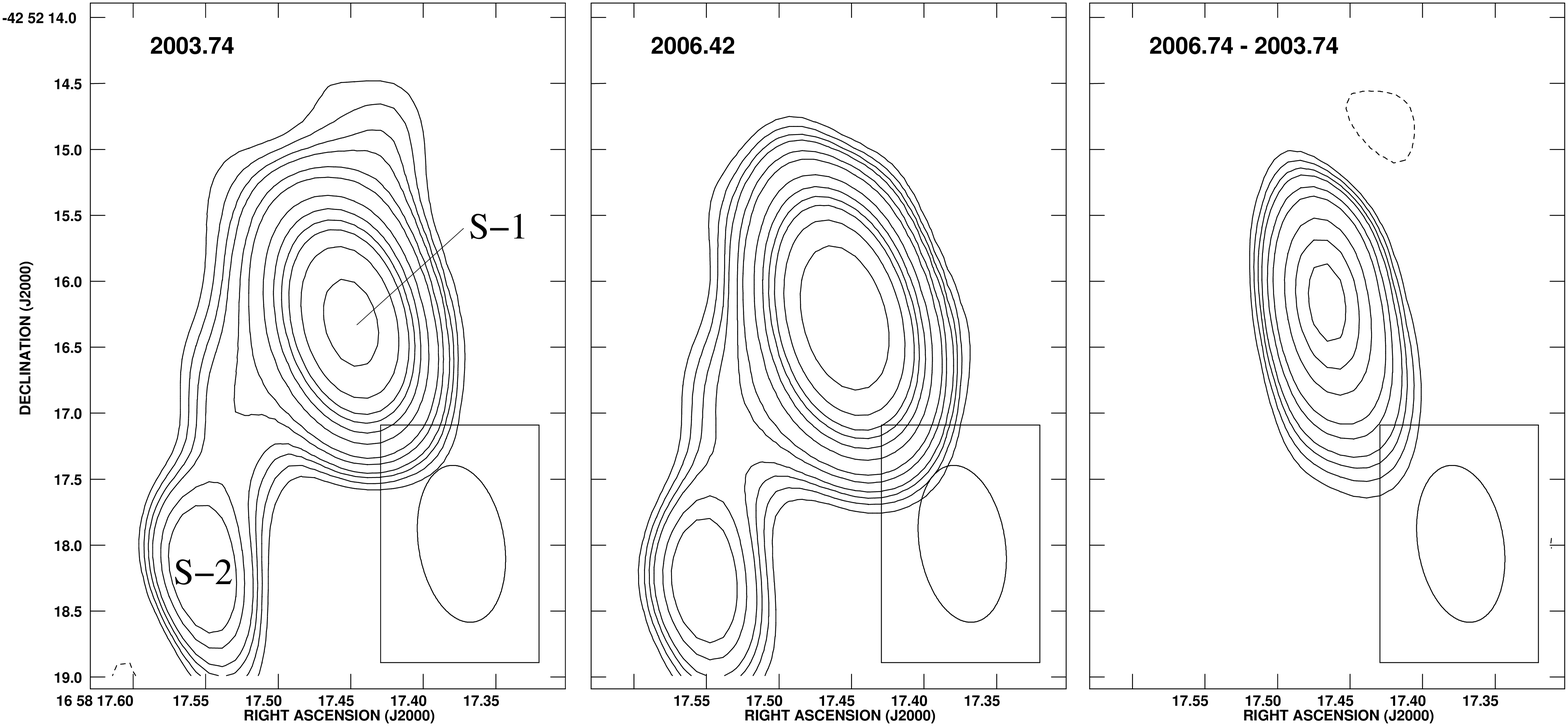}
\end{center}
\caption{VLA contour images at 8.46 GHz towards
the component S-1 in
IRAS~16547-4247 for epochs 2003.74 (left) and 2006.42 (center),
as well as the difference image (2006.42 - 2003.74).
Contours are -25, -20, -15, -10, -8, -6, -5, -4, 4, 5,
6, 8, 10, 15, 20, 25, 30, 40, 60, 100, 140, 160, and 200 times
27 $\mu$Jy beam$^{-1}$ for the 2003.74 and 2006.42 images
and 38 $\mu$Jy beam$^{-1}$ for the difference image.
The half power contour of the synthesized beams
($1\rlap.{''}20 \times 0\rlap.{''}65$; PA = $9^\circ$)
is shown in the bottom right corner.
Individual sources are identified
in the 2003.74 image.
\label{fig6}}
\end{figure}

\clearpage

\begin{figure}
\begin{center}
\includegraphics[scale=0.75, angle=0]{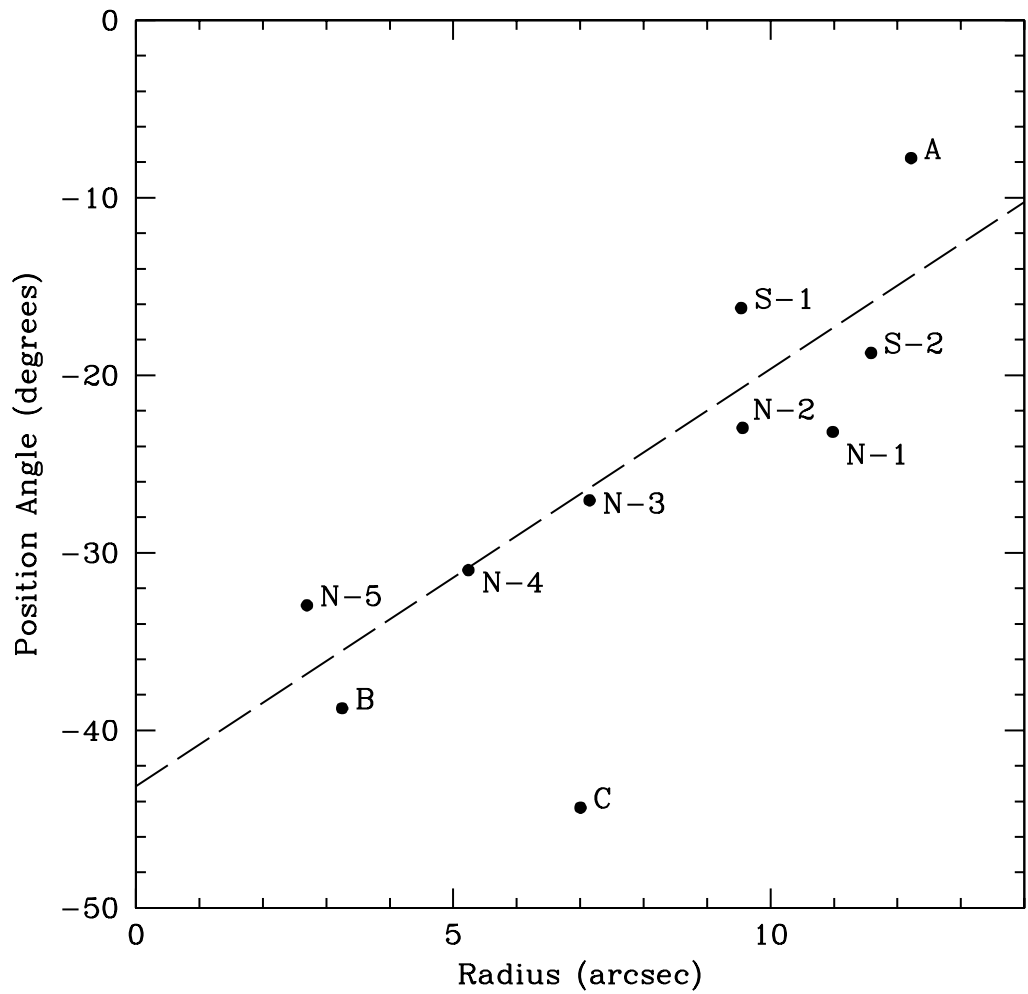}
\end{center}
\caption{Position angle of the jet components as a function
of radial offset from the jet center.
The components south of the jet have been folded by 180$^\circ$
in this figure.
The dashed line is the linear least squares fit to the components,
with the exception of source C, that is taken to be an
independent star.
\label{fig7}}
\end{figure}

\begin{figure}
\begin{center}
\includegraphics[scale=0.75, angle=0]{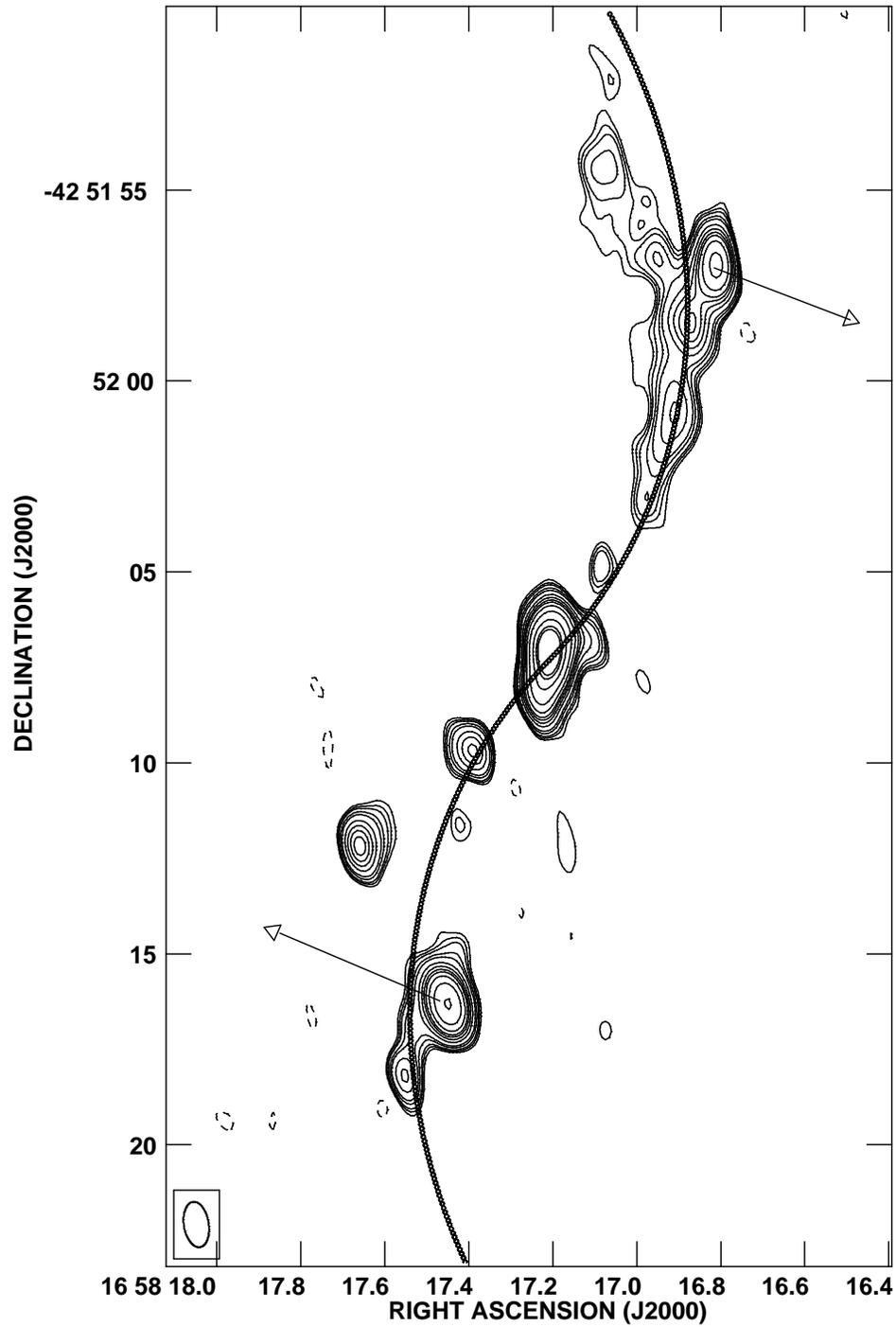}
\end{center}
\caption{VLA contour image at 8.46 GHz towards 
IRAS~16547-4247 for epoch 2003.74. 
Contours and beam are as in Figure 2.
The solid line indicates the position of the spiral model
discussed in the text. The arrows indicate the proper motions
of components N-1 and S-1 for a period of 300 years.
\label{fig8}}
\end{figure}

\begin{figure}
\begin{center}
\includegraphics[scale=0.75, angle=0]{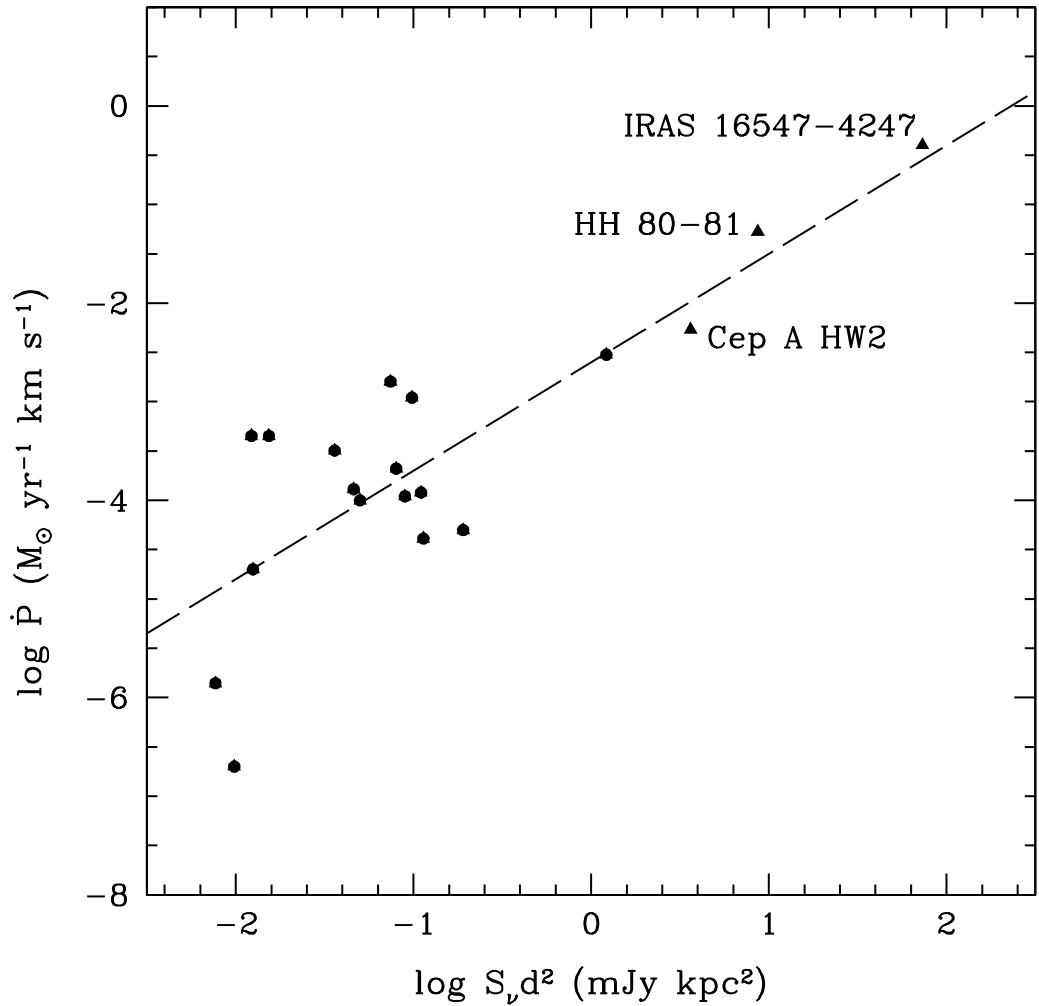}
\end{center}
\caption{Momentum rate in the molecular outflow, $\dot P$,
versus the radio flux density times distance squared,
$S_\nu d^2$. The solid dots are the 16 low mass young stars from
the study of Anglada et al.~(1992). We have included three
high mass young stars (solid triangles, see also labels
in figure) from this paper as well as from
the references given in Table 3. The dashed line is the fit of
Anglada et al.~(1992) to the low mass young stars. The
high mass young stars fall reasonably well on the correlation,
suggesting a common nature for thermal jets associated with
low and high mass young stars.
\label{fig9}}
\end{figure}

\clearpage
\begin{table}
\small
%
\caption{Observed 3.6 cm parameters of sources in the field \label{tab1}}
\begin{tabular}{@{}lllll}
\br
Component & $\alpha$(2000)$^{\rm a}$ & $\delta$(2000)$^{\rm a}$ & S$_\nu^{\rm b}$ &  \cr
\ns
(Epoch) & ($16^h~58^m$) & ($-42^\circ$) & (mJy) & Deconvolved Size$^{\rm c}$ \cr 
\mr
A(2003)  & 17$\rlap.^s$0597$\pm$0.0077 & 51$'$ 55$\rlap.{''}$039$\pm$0$\rlap.{''}$240 &
     1.58$\pm$0.23 &
     4\rlap.{''}2$\pm$0\rlap.{''}6$\times$1\rlap.{''}2$\pm$0\rlap.{''}2;
     9$^\circ \pm$4$^\circ$ \\
A(2006)  & 17$\rlap.^s$0641$\pm$0.0072 & 51$'$ 55$\rlap.{''}$212$\pm$0$\rlap.{''}$205 &
     1.60$\pm$0.20 &
     4\rlap.{''}1$\pm$0\rlap.{''}5$\times$1\rlap.{''}5$\pm$0\rlap.{''}2;
     4$^\circ \pm$5$^\circ$ \\
\hline
N-1(2003) & 16$\rlap.^s$8165$\pm$0.0005 & 51$'$ 57$\rlap.{''}$048$\pm$0$\rlap.{''}$011 &
     $2.86\pm0.07$  & 
     1\rlap.{''}06$\pm$0\rlap.{''}04$\times$0\rlap.{''}31$\pm$0\rlap.{''}05;
     163$^\circ \pm$3$^\circ$ \\
N-1(2006) & 16$\rlap.^s$8136$\pm$0.0004 & 51$'$ 57$\rlap.{''}$060$\pm$0$\rlap.{''}$009 &
     $2.85\pm0.06$  &
     1\rlap.{''}15$\pm$0\rlap.{''}03$\times$0\rlap.{''}23$\pm$0\rlap.{''}08;
     156$^\circ \pm$2$^\circ$ \\
\hline
N-2(2003) & 16$\rlap.^s$8706$\pm$0.0015 & 51$'$ 58$\rlap.{''}$339$\pm$0$\rlap.{''}$028 &
    $2.35\pm0.10$  & 
    1\rlap.{''}65$\pm$0\rlap.{''}09$\times$0\rlap.{''}80$\pm$0\rlap.{''}08;
    162$^\circ \pm$5$^\circ$ \\
N-2(2006) & 16$\rlap.^s$8669$\pm$0.0011 & 51$'$ 58$\rlap.{''}$249$\pm$0$\rlap.{''}$019 &
    $2.24\pm0.07$  &
    1\rlap.{''}47$\pm$0\rlap.{''}07$\times$0\rlap.{''}76$\pm$0\rlap.{''}07;
    151$^\circ \pm$5$^\circ$ \\
\hline
N-3(2003) & 16$\rlap.^s$9142$\pm$0.0021 & 52$'$ 00$\rlap.{''}$776$\pm$0$\rlap.{''}$061 &
    $1.82\pm0.12$ 
    & 2\rlap.{''}52$\pm$0\rlap.{''}17$\times$0\rlap.{''}62$\pm$0\rlap.{''}10;
    168$^\circ \pm$3$^\circ$ \\
N-3(2006) & 16$\rlap.^s$9114$\pm$0.0022 & 52$'$ 00$\rlap.{''}$921$\pm$0$\rlap.{''}$052 &
    $1.76\pm0.10$
    & 2\rlap.{''}52$\pm$0\rlap.{''}14$\times$0\rlap.{''}78$\pm$0\rlap.{''}09;
    162$^\circ \pm$3$^\circ$ \\
\hline
E(2003)  & 18$\rlap.^s$2348$\pm$0.0080 & 52$'$ 02$\rlap.{''}$183$\pm$0$\rlap.{''}$115 &
     0.16$\pm$0.05 &
     $ \leq 1\rlap.{''}2^{\rm d}$ \\
E(2006)  & 18$\rlap.^s$2128$\pm$0.0044 & 52$'$ 02$\rlap.{''}$319$\pm$0$\rlap.{''}$106 &
     0.12$\pm$0.04 &
     $ \leq 1\rlap.{''}1^{\rm d}$ \\
\hline
N-4(2003) & 16$\rlap.^s$9645$\pm$0.0061 & 52$'$ 02$\rlap.{''}$650$\pm$0$\rlap.{''}$164 &
    $0.87\pm0.14$
    & 2\rlap.{''}66$\pm$0\rlap.{''}43$\times$0\rlap.{''}88$\pm$0\rlap.{''}27;
    171$^\circ \pm$9$^\circ$ \\
N-4(2006) & 16$\rlap.^s$9851$\pm$0.0033 & 52$'$ 03$\rlap.{''}$286$\pm$0$\rlap.{''}$092 &
    $0.86\pm0.09$
    & 2\rlap.{''}40$\pm$0\rlap.{''}25$\times$0\rlap.{''}71$\pm$0\rlap.{''}16;
    170$^\circ \pm$5$^\circ$ \\
\hline
N-5(2003) & 17$\rlap.^s$0764$\pm$0.0053 & 52$'$ 04$\rlap.{''}$881$\pm$0$\rlap.{''}$079 &
    $0.27\pm0.06$
    & $\leq 1\rlap.{''}0^{\rm d}$ \\
N-5(2006) & 17$\rlap.^s$0715$\pm$0.0024 & 52$'$ 04$\rlap.{''}$737$\pm$0$\rlap.{''}$046 &
    $0.30\pm0.04$
    & $\leq 0\rlap.{''}7^{\rm d}$ \\
\hline
D(2003)  & 17$\rlap.^s$1265$\pm$0.0016 & 52$'$ 06$\rlap.{''}$840$\pm$0$\rlap.{''}$031 &
     0.50$\pm$0.03 &
     $ \leq 0\rlap.{''}7^{\rm d}$ \\
D(2006)  & 17$\rlap.^s$1211$\pm$0.0012 & 52$'$ 06$\rlap.{''}$833$\pm$0$\rlap.{''}$023 &
     0.48$\pm$0.02 &
     $ \leq 0\rlap.{''}6^{\rm d}$ \\
\hline
Jet(2003)  & 17$\rlap.^s$2097$\pm$0.0001 & 52$'$ 07$\rlap.{''}$142$\pm$0$\rlap.{''}$003 &
     10.4$\pm$0.1 & 
     1\rlap.{''}21$\pm$0\rlap.{''}01$\times$0\rlap.{''}15$\pm$0\rlap.{''}03; 
     165$^\circ \pm$1$^\circ$ \\
Jet(2006)  & 17$\rlap.^s$2093$\pm$0.0001 & 52$'$ 07$\rlap.{''}$150$\pm$0$\rlap.{''}$002 &
     11.3$\pm$0.1 & 
     1\rlap.{''}20$\pm$0\rlap.{''}01$\times$0\rlap.{''}14$\pm$0\rlap.{''}02; 
     163$^\circ \pm$1$^\circ$ \\
\hline
B(2003)  & 17$\rlap.^s$3946$\pm$0.0013 & 52$'$ 09$\rlap.{''}$675$\pm$0$\rlap.{''}$019 &
     0.81$\pm$0.06 &
    $ \leq 0\rlap.{''}6^{\rm d}$ \\
B(2006)  & 17$\rlap.^s$3904$\pm$0.0009 & 52$'$ 09$\rlap.{''}$732$\pm$0$\rlap.{''}$015 &
     0.85$\pm$0.04 &
    $ \leq 0\rlap.{''}5^{\rm d}$ \\
\hline
C(2003)  & 17$\rlap.^s$6549$\pm$0.0010 & 52$'$ 12$\rlap.{''}$150$\pm$0$\rlap.{''}$018 &
     1.10$\pm$0.06 &
    $ \leq 0\rlap.{''}6^{\rm d}$ \\
C(2006)  & 17$\rlap.^s$6603$\pm$0.0007 & 52$'$ 12$\rlap.{''}$221$\pm$0$\rlap.{''}$012 &
     0.97$\pm$0.04 &
    $ \leq 0\rlap.{''}5^{\rm d}$ \\
\hline
S-1(2003) & 17$\rlap.^s$4519$\pm$0.0004 & 52$'$ 16$\rlap.{''}$301$\pm$0$\rlap.{''}$006 &
    $3.65\pm0.06$  & 
    0\rlap.{''}59$\pm$0\rlap.{''}04$\times$0\rlap.{''}40$\pm$0\rlap.{''}05;
    51$^\circ \pm$9$^\circ$ \\
S-1(2006) & 17$\rlap.^s$4556$\pm$0.0002 & 52$'$ 16$\rlap.{''}$284$\pm$0$\rlap.{''}$003 &
    $4.97\pm0.05$  &
    0\rlap.{''}66$\pm$0\rlap.{''}02$\times$0\rlap.{''}38$\pm$0\rlap.{''}02;
    35$^\circ \pm$4$^\circ$ \\
\hline
S-2(2003) & 17$\rlap.^s$5481$\pm$0.0013 & 52$'$ 18$\rlap.{''}$113$\pm$0$\rlap.{''}$031 &
    0.63$\pm$0.06 &
    $ \leq 0\rlap.{''}7^{\rm d}$ \\ 
S-2(2006) & 17$\rlap.^s$5473$\pm$0.0011 & 52$'$ 18$\rlap.{''}$210$\pm$0$\rlap.{''}$026 &
    0.69$\pm$0.05 &
    $ \leq 0\rlap.{''}8^{\rm d}$ \\

\br
\end{tabular}
$^{\rm a}$ Peak position. Right ascension ($\alpha$) given in hours, 
minutes, and seconds, and declination ($\delta$), given in degrees, arcmins, 
and arcsecs. The errors given are formal statistical errors. The absolute 
positional error of the images are estimated to be 0$\rlap .{''}$2
in right ascension and 0$\rlap .{''}$5 in declination.
The sources are listed in order of decreasing declination.

$^{\rm b}$ Total flux density. The errors given are formal statistical errors.
The absolute flux density error is estimated to be on the order of 10\%.

$^{\rm c}$ Deconvolved dimensions of the source: FWHM major axis $\times$
FWHM minor axis; position angle of major axis.

$^{\rm d}$ Unresolved
\end{table}
\normalsize

\clearpage

\begin{table}
\small
%
\caption{Proper Motions of Sources \label{tab2}}
\begin{tabular}{@{}lccccccccc}
\br
 & D$^{\rm a}$ & PA(O)$^{\rm b}$ & PA(P)$^{\rm c}$ & $\mu(\alpha)^{\rm d}$ & $\mu (\delta)^{\rm e}$ &
$\mu^{\rm f}$ & $PA(\mu)^{\rm g}$ & $\mu (R)^{\rm h}$ & $\mu (T)^{\rm i}$ \cr
\ns
Source & ({''}) & ($^\circ$) & ($^\circ$) & (mas~yr$^{-1}$) & (mas~yr$^{-1}$) &
(mas~yr$^{-1}$) & ($^\circ$) & (mas~yr$^{-1}$) & (mas~yr$^{-1}$)  \cr 
\mr

 A & 12.215 & -7.76  & -15 &  18$\pm$43 &  -65$\pm$118  &   67$\pm$114 & 
 164$\pm$97 &  -67$\pm$114 &  0$\pm$53  \\
 N-1  & 10.981  & -23.19  &  -18 &  -12$\pm$3 &  -4$\pm$5 & 13$\pm$3  &
 -111$\pm$14  &  -1$\pm$5 &  -13$\pm$3 \\
 N-2 &  9.560  &  -22.96  & -21 &  -15$\pm$8  & 34$\pm$13  &  37$\pm$12 &
 -24$\pm$19 & 36$\pm$12 &   -5$\pm$8 \\
 N-3   & 7.147 &  -27.04  &  -27  &  -11$\pm$12 &  -54$\pm$30  & 55$\pm$29  &
 -168$\pm$30 &  -49$\pm$29 &   -26$\pm$15  \\
 E & 12.316  & 66.26 &   -15 &   -90$\pm$37  &  -51$\pm$58  & 104$\pm$43 &
 -119$\pm$24 &  -24$\pm$57  &  -101$\pm$39 \\
 N-4 &  5.239 &  -30.98 &   -31 & 85$\pm$28 &  -237$\pm$70 & 252$\pm$67  &
 160$\pm$15 &   -251$\pm$68  & 16$\pm$34  \\
 N-5  & 2.695 &  -32.96  &  -37 &  -20$\pm$24  &  54$\pm$34  & 57$\pm$33  &
 -21$\pm$33 &  57$\pm$33  &  -5$\pm$25  \\
 D  &  0.964  &  -71.73  &  -41 &  -22$\pm$8 & 3$\pm$14  &  22$\pm$8   &
 -83$\pm$21 &   9$\pm$14 &   -21$\pm$9 \\
  Jet  & 0.000  &   0.00  &   -43 &    -2$\pm$1  &  -3$\pm$1  &   3$\pm$1  &
 -151$\pm$20 &  -2$\pm$1  &  -2$\pm$1 \\
  B &  3.248 &   141.24  &    144 &  -17$\pm$6 &   -21$\pm$9   &  27$\pm$8 &
 -141$\pm$17 &   17$\pm$9  &   22$\pm$7  \\
  C  &  7.004 &  135.65  &  153 &   22$\pm$5  &  -26$\pm$8  & 35$\pm$7  &
  140$\pm$12  & 31$\pm$8  &   -16$\pm$5  \\
 S-1  & 9.538 &  163.79  &  159 &  15$\pm$2  &  6$\pm$3 & 16$\pm$2 & 
 67$\pm$7 &  -3$\pm$2 &   -16$\pm$2  \\
 S-2   & 11.585  &  161.26 &  164 &  -3$\pm$7  &  -36$\pm$15  &  36$\pm$15  &
 -175$\pm$24 &  35$\pm$15  &  11$\pm$8  \\

\br
\end{tabular}
$^{\rm a}$ Displacement with respect to the jet in arcsec.

$^{\rm b}$ Observed position angle with respect to the jet in degrees.

$^{\rm c}$ Predicted position angle in the precession model.

$^{\rm d}$ Right Ascension proper motion in milliarcsec per year. (1 mas yr$^{-1}$ = 14 km s$^{-1}$ at a distance of 2.9 Kpc)

$^{\rm e}$ Declination proper motion in milliarcsec per year.

$^{\rm f}$ Total proper motion in milliarcsec per year.

$^{\rm g}$ Position angle of the total proper motion in degrees. 

$^{\rm h}$ Radial proper motion with respect to an axis with PA = $-16^\circ$,
in milliarcsec per year. A negative sign corresponds to motions approaching the jet. 

$^{\rm i}$ Transversal proper motion with respect to an axis with PA = $-16^\circ$, in milliarcsec per year. 
A positive sign corresponds to counterclockwise motion.

$^{\rm j}$ Note that if the maps were aligned to make the position of the jet fixed, the 
proper motions of the other components would change by 2 and 3 mas yr$^{-1}$ in Right Ascension and Declination, respectively.

\end{table}
\normalsize

\clearpage

\small
\begin{table}
\caption{Jet and Molecular Outflow Parameters of Three Massive Young Stars.\label{tbl-3}}
\begin{indented}
\item[]\begin{tabular}{@{}lllll}
\br
 &  Distance & $S_\nu$(3.6 cm) & $\dot P$ & \\
\ns
Source &  (kpc) & (mJy) & ($M_\odot$ yr$^{-1}$ km s$^{-1}$) & References$^{\rm a}$ \\
\mr
IRAS~16547-4247 & 2.9 & 8.7 & $4.0\times10^{-1}$ & 1,2,3,4 \\ 
HH~80-81 & 1.7 & 3.0 & $5.3\times10^{-2}$ & 5,6  \\
Cep A HW2 & 0.725 & 6.9 & $5.4\times10^{-3}$ & 7,8 \\
\br
\end{tabular}
\item[] $^{\rm a}$ 1) Garay et al. (2007); 2) Garay et al. (2003);
3) Rodr\'\i guez et al. (2005); 4) this paper; 5) Mart\'\i\ et al. (1993);
6) Yamashita et al. (1989); 7) Curiel et al. (2006); 8) Narayanan \& Walker (1996).
\end{indented}
\end{table}
\normalsize

\clearpage

\end{document}